\documentclass[prb,showpacs,twocolumn,preprintnumbers,amsmath,groupedaddress,amssymb,superscriptaddress,floatfix]{revtex4}
\usepackage[dvips]{graphicx}
\usepackage[latin1]{inputenc}
\usepackage{xcolor,bm}
\usepackage{amsmath,multirow}
\usepackage{nicefrac}
\begin{document}
\draft

\hyphenation{a-long}

\title{Amorphous ferromagnetism and re-entrant magnetic glassiness in Sm$_{2}$Mo$_{2}$O$_{7}$: \\new insights into the electronic phase diagram of pyrochlore molybdates}

\author{G.~Prando}\email[E-mail: ]{g.prando@ifw-dresden.de}\affiliation{Leibniz-Institut f\"ur Festk\"orper- und Werkstoffforschung (IFW) Dresden, D-01171 Dresden, Germany}
\author{P.~Carretta}\affiliation{Dipartimento di Fisica and Unit\`a CNISM di Pavia, Universit\`a di Pavia, I-27100 Pavia, Italy}
\author{A.~U.~B.~Wolter}\affiliation{Leibniz-Institut f\"ur Festk\"orper- und Werkstoffforschung (IFW) Dresden, D-01171 Dresden, Germany}
\author{R.~Saint-Martin}\affiliation{Laboratoire de Physico-Chimie de l'Etat Solide, ICMMO, UMR8182, Université Paris-Sud, F-91405 Orsay, France}
\author{A.~Revcolevschi}\affiliation{Laboratoire de Physico-Chimie de l'Etat Solide, ICMMO, UMR8182, Université Paris-Sud, F-91405 Orsay, France}
\author{B.~B\"uchner}\affiliation{Leibniz-Institut f\"ur Festk\"orper- und Werkstoffforschung (IFW) Dresden, D-01171 Dresden, Germany}\affiliation{Institut f\"ur Festk\"orperphysik, Technische Universit\"at Dresden, D-01062 Dresden, Germany}

\widetext

\begin{abstract}
We discuss the magnetic properties of a Sm$_{2}$Mo$_{2}$O$_{7}$ single crystal as investigated by means of different experimental techniques. In the literature, a conventional itinerant ferromagnetic state is reported for the Mo$^{4+}$ sublattice below $\sim 78$ K. However, our results of dc magnetometry, muon spin spectroscopy ($\mu^{+}$SR) and high-harmonics magnetic ac susceptibility unambiguously evidence highly disordered conditions in this phase, in spite of the crystalline and chemical order. This disordered magnetic state shares several common features with amorphous ferromagnetic alloys. This scenario for Sm$_{2}$Mo$_{2}$O$_{7}$ is supported by the anomalously high values of the critical exponents, as mainly deduced by a scaling analysis of our dc magnetization data and confirmed by the other techniques. Moreover, $\mu^{+}$SR detects a significant static magnetic disorder at the microscopic scale. At the same time, the critical divergence of the third-harmonic component of the ac magnetic susceptibility around $\sim 78$ K leads to additional evidence towards the glassy nature of this magnetic phase. Finally, the longitudinal relaxation of $\mu^{+}$ spin polarization (also supported by results of ac susceptibility) evidences re-entrant glassy features similar to amorphous ferromagnets.
\end{abstract}


\date{\today}

\maketitle

\narrowtext

\section{Introduction}

Pyrochlore molybdates $R_{2}$Mo$_{2}$O$_{7}$ have attracted remarkable attention in recent years due to the wealth of electronic ground states and physical properties they display as a function of external (e.~g., pressure)\cite{Ape06,Mir06,Igu09,Gar10} and internal [e.~g., substitution of rare-earth ($R$) ions]\cite{Kat00,Mor01,Kez06,Han07,Ehl10,Gar10,Igu11,Ued12} parameters. A Mott-like metal-to-insulator transition\cite{Ima98} (MIT) is observed by modifying the ionic radius of the $R$ ion, from a ferromagnetic metallic (FMM) to a spin-glass insulating (SGI) phase,\cite{Kat00,Mor01,Kez06} even if glassy features are detected also in the FMM phase very close to the MIT boundary.\cite{Kat00} Compounds with $R$ = Gd, Sm, Nd are reported to belong to the FMM phase while, e.~g., Y$_{2}$Mo$_{2}$O$_{7}$ clearly displays SGI behaviour similarly to the cases $R$ = Lu, Yb, Er, Ho, Dy, Tb.\cite{Mor01,Tag99,Han07,Sin12,Cla13,Cla14} Interestingly, the origin of the disorder leading to magnetic glassiness in Y$_{2}$Mo$_{2}$O$_{7}$ is still controversial and object of intensive investigations. Glassy features are indeed clearly detected even in the presence of negligible degrees of chemical disorder,\cite{Dun96,Gar99,Ker01,Sag05,Ofe10,Sil14} a property seemingly hallmark of several other magnetic pyrochlores.\cite{Lag14} From a local point of view, bond-randomness is accepted to be the origin of glassiness in Y$_{2}$Mo$_{2}$O$_{7}$,\cite{Boo00} the principal source of disorder arising from Y-Mo rather than Mo-Mo pairs.\cite{Gre09}

However, more recent works\cite{Sol03,Mot10,Mot10b,Shi13,Sil14} suggest that a complicated interplay between spin and orbital degrees of freedom\cite{Wit14} on the basis of a Kugel-Khomskii mechanism\cite{Kug82} may be crucial in determining the overall magnetic properties of pyrochlore molybdates. In particular, on the basis of a multi-orbital Hubbard model, Shinaoka et al.\cite{Shi13} conclude that Y$_{2}$Mo$_{2}$O$_{7}$ should be considered as the realization of a ``\textit{spin-orbital frustrated Mott insulator}'' rather than a conventional frustrated magnet. Possibly, the main reason for the complicated wealth of electronic and magnetic ground states (GS) in pyrochlore molybdates is then, on one side, the competition between Coulombic energy and spin-orbit coupling (both of them being of comparable intensity in $4d$ systems).\cite{Shi13}

On the other hand, the peculiar geometrical properties introduced by the pyrochlore lattice are also expected to play an important role.\cite{Sub83,Ram94,Gar10} More generally, $A_{2}^{3+} B_{2}^{4+}$O$_{7}$ compounds typically display such crystalline structure characterized by a corner-sharing arrangement of tetrahedra identical for the two interpenetrating sublattices of $A$ and $B$ ions. An extremely rich variety of magnetic phases is observed for different $A$ and $B$ ions, where typically $A$ is a $R$ ion while $B$ is chosen among transition metal ($Tm$) elements.\cite{Sub83,Gar10} The most striking effect of the peculiar geometrical properties of the pyrochlore lattice is likely the emergence of the spin-ice phase for the insulating compounds with $R$ = Ho, Dy and $Tm$ = Sn, Ti.\cite{Har97,Moe98,Bra01n,Pra09,Zho12,Cas12} Here, the single-ion easy-axis anisotropy of $R$ ions along the local $\langle 1 \; 1 \; 1 \rangle$ crystallographic directions\cite{Moe98,Ros00} and the mainly dipolar magnetic interactions\cite{Her00,Bra01,Isa04} allow for a locally ordered two-in/two-out arrangement of magnetic moments in the GS. However, the GS is highly frustrated on the macroscopic level and geometrically equivalent to the disorder of the common $I_{h}$ water ice.\cite{Pau35,Gia36,Ram99,XKe07,Pra09,Zho12} These particular features of the magnetic GS in spin-ice materials have triggered an enormous interest in the recent years since emergent magnetic excitations\cite{Lee02,Fan03,Bal10} proper of the system can be described in terms of magnetic monopoles.\cite{Ryz05,Cas08,Jau09} The detection of experimental signatures proper of magnetic monopoles has been recently claimed in spin-ices by means of both macroscopic\cite{Mor09,Gib10,Cas11,Bov13,Pau14} and local magnetic techniques,\cite{Bra09,Sal12} even if the topic is still highly controversial and debated particularly in the case of muon spin spectroscopy ($\mu^{+}$SR)  results.\cite{Blu12}

Recently, the detection of spin-ice-like phases has been claimed also for the metallic molybdate pyrochlore Sm$_{2}$Mo$_{2}$O$_{7}$\cite{Sin08} and for other metallic Ir-based pyrochlores.\cite{Nak06,Mac07,Ike08,Tok14} In the particular case of Sm$_{2}$Mo$_{2}$O$_{7}$, based on macroscopic measurements (in particular, specific heat), the material has been reported to sustain an ``\textit{ordered}'' spin-ice phase at low temperatures, where the ordering (polarizing) effect on Sm$^{3+}$ magnetic moments would come from the internal molecular field generated by the Mo$^{4+}$ sublattice.\cite{Sin08} However, differently from the case of dipolar spin-ice materials,\cite{Ros00} direct information on the crystal-field-split electronic levels of Sm$^{3+}$ cannot be accessed easily by neutron techniques in view of the high absorption cross section of Sm$^{3+}$ ions.\cite{Rya09,Gar10} This topic is still highly controversial for Sm$_{2}$Mo$_{2}$O$_{7}$. Local easy-axis magnetic anisotropy along $\langle 1 \; 1 \; 1 \rangle$ directions is claimed in Ref.~\onlinecite{Sin08}. At the same time, from theoretical arguments about crystal field potential, Gardner et al.\cite{Gar10} report a more likely easy-plane configuration for Sm$^{3+}$ magnetic moments, similarly to the case of Er$^{3+}$ magnetic moments in a pyrochlore lattice.\cite{Sid99,Cha03,Lag05,Dal12} More generally, it should be also stressed that, while the magneto-transport properties of Sm$_{2}$Mo$_{2}$O$_{7}$ are well-characterized in view of the anomalous Hall effect arising in Sm- and Nd-based pyrochlore molybdates,\cite{Tag99,Kat00,Tag01,Yas01,Tag03} its microscopic magnetic properties are -- to the best of our knowledge -- still mainly unexplored. A $\mu^{+}$SR study on a single crystal of Sm$_{2}$Mo$_{2}$O$_{7}$ has been reported in the past.\cite{Jo05} However, the instrumental time ($t$) resolution did not allow to investigate the local magnetic features in detail. Moreover, characteristic transition temperatures reported in Ref.~\onlinecite{Jo05} are quite low compared to other reports in the literature, pointing towards a strong effect of O$^{2-}$ vacancies on the overall properties of the material, a well-known problem associated with single crystals of pyrochlore molybdates.\cite{Gar10}

In this work, we report on a detailed investigation of a high-quality single crystal of Sm$_{2}$Mo$_{2}$O$_{7}$ as performed by means of different experimental techniques (dc magnetometry, $\mu^{+}$SR, high-harmonics ac susceptibility). The magnetic phase of the Mo$^{4+}$ sublattice develops for $T < T_{C} \simeq 78$ K, in agreement with several previous reports in the literature. The $T_{C}$ value would be significantly reduced by a substantial amount of O$^{2-}$ vacancies,\cite{Gar10,Sil14,Cao95} showing that this issue can be safely neglected for the currently investigated sample. Such phase for Mo$^{4+}$ is typically discussed in the literature as a conventional itinerant ferromagnetic state. However, our results clearly detect a complicated superposition of conventional and highly disordered magnetic behaviours below $\sim 78$ K sharing several common features with amorphous ferromagnetic alloys (AmFA) and with other SGI pyrochlore molybdates. This scenario for Sm$_{2}$Mo$_{2}$O$_{7}$ is supported by the anomalously high values deduced for the critical exponents of the magnetic transition, approaching values typically reported for AmFA. These were calculated by a scaling-analysis of the dc magnetization data and confirmed by $\mu^{+}$SR and first-harmonic ac susceptibility. At the same time, $\mu^{+}$SR detects a sizeable static magnetic disorder at the microscopic scale resulting in strongly damped coherent oscillations in the $t$-depolarization of the $\mu^{+}$ spin. Moreover, the critical divergence of the third-harmonic component of the magnetic ac susceptibility around $\sim 80$ K leads to additional evidence towards the disordered nature of this magnetic phase. Some degree of magnetic glassiness has been reported in the literature also in the FMM phase near to the MIT boundary.\cite{Kat00} However, Sm$_{2}$Mo$_{2}$O$_{7}$ is located far enough from such boundary and glassy features are typically neglected in this case.\cite{Igu09} Finally, as typical for several amorphous ferromagnets, a re-entrant spin-glass phase is evidenced at low temperatures by means of both the longitudinal magnetic relaxation of $\mu^{+}$ and by magnetic ac susceptibility. Accordingly, our results shed new light on the magnetic properties of Sm$_{2}$Mo$_{2}$O$_{7}$ and on the overall electronic phase diagram commonly accepted for pyrochlore molybdates, which is proposed in a new version at the end of this paper.

\section{Experimentals}

\subsection{Single crystal growth}

The single crystal of Sm$_{2}$Mo$_{2}$O$_{7}$ was grown using the optical floating-zone method in a purified Ar atmosphere. The successful growth of cm$^{3}$-size crystals of Sm$_{2}$Mo$_{2}$O$_{7}$ is achieved by overcoming specific difficulties, including the decomposition of the pyrochlore phase at low temperatures, the highly volatile nature of MoO$_{2}$, and the dependence of the oxidation state of Mo on small variations in the growth atmosphere. Further details about the growth procedure can be found in Ref.~\onlinecite{Sin07}.

\subsection{dc magnetometry}\label{SectionExperimentalsDcMagnetometry}

dc magnetometry measurements were performed by using the commercial magnetometers Magnetic Property Measurement System MPMS-XL7 and Physical Property Measurement System PPMS (Quantum Design). The dc magnetization ($M_{dc}$) was measured in static polarizing magnetic fields up to $H_{0} = 70$ kOe as a function of both temperature ($T$) and $H_{0}$.

Although the shape of the investigated single crystal is not regular, it can approximately be modelled as an elliptic disk.\cite{Sin07} With reference to the notation reported in Ref. \onlinecite{Bel05}, one has $2a = 4.2 \pm 0.1$ mm, $2b = 2.9 \pm 0.1$ mm for the two axes of the ellipse and $t = 1.5 \pm 0.1$ mm for its height. The elliptic faces are characterized by the Miller indexes  $(1 \; 1 \; 1)$. For all the experiments presented in this paper, the magnetic field was applied perpendicular to the elliptic faces, namely ${\bf H}_{0} \parallel [1 \; 1 \; 1]$ (within an accuracy of $\pm \; 1°$). Under these circumstances, the demagnetization factor is estimated as $D \simeq 0.45$.\cite{Bel05} In the following, the dc magnetization is expressed in molar units (per mole of formula units, f.u.) and the value $D_{m} = \left(N_{mol}/V\right) \times D$ is used to account for demagnetization effects, accordingly. Here, the pre-factor accounts for the molar density of the material, equal to $\sim 7.5 \times 10^{-3}$ mol cm$^{-3}$, with $N_{mol}$ as the number of moles in the volume $V$.

The non-negligible value of $D$ makes it necessary to properly take demagnetization effects into account before the data analysis. In particular, concerning the $M_{dc}$ vs. $H_{0}$ measurements to be discussed later in section \ref{SectionResultsDcMagnetization}, the magnetic field value must be
corrected as\cite{Seg95}
\begin{equation}\label{EqDemagFactMvsH}
H_{i} = H_{a} - 4 \pi D_{m} M_{dc}
\end{equation}
where $H_{a}$ and $H_{i}$ are the applied and effective (intrinsic) magnetic fields, respectively. In the case of $M_{dc}$ vs. $T$ scans at vanishing values of the magnetic field, one has $\chi_{m} = M_{dc}/H_{a}$ for the measured susceptibility and
\begin{equation}\label{EqDemagFactMvsT}
\frac{1}{\chi_{i}} = \frac{1}{\chi_{m}} - 4 \pi D_{m}
\end{equation}
for the intrinsic susceptibility $\chi_{i} = M_{dc}/H_{i}$.

Only intrinsic data of both magnetic field and susceptibility are presented in this paper, unless explicitly stated otherwise. From now on, the indexes $i$ are dropped for the aim of clarity.

\subsection{Muon spin spectroscopy}\label{SectionExperimentalsMuSR}
The Sm$_{2}$Mo$_{2}$O$_{7}$ single crystal was investigated by means of $\mu^{+}$SR at the S$\mu$S muon source of the Paul Scherrer Institut, Switzerland (GPS spectrometer, $\pi$M$3$ beamline). Measurements were performed in the temperature range $T = 1.6 - 120$ K and in conditions either of zero magnetic field (ZF-$\mu^{+}$SR) or of longitudinal static magnetic field (LF-$\mu^{+}$SR) up to $H = 2000$ Oe applied parallel to the $[1 \; 1 \; 1]$ crystallographic axis of the crystal.

The main output of a $\mu^{+}$SR experiment is the so-called $t$ dependence of the decay asymmetry $A_{T}(t)$ for $\mu^{+}$.\cite{Blu99,Yao11} In general, this quantity can be directly related to the $t$ dependence of the spin polarization $P_{T}(t)$ of the statistical ensemble of $\mu^{+}$ as subject to local magnetic field in the sample, namely\cite{Blu99,Yao11}
\begin{equation}\label{EqGeneralFittingZF}
P_{T}(t) \equiv \frac{A_{T}(t)}{A_{0}} = G_{T}^{s}(t),
\end{equation}
where $A_{0}$ is an instrumental parameter quantifying the maximum amplitude of the signal. The function $G_{T}^{s}(t)$ is connected to the magnetic features of the investigated material, and its particular form is discussed in detail later in Sect.~\ref{SectionResultsMuSR}.

\subsection{Magnetic ac susceptibility}\label{SectionExperimentalsAcSusceptibility}

Measurements of ac magnetic susceptibility were performed by means of a commercial susceptometer Physical Property Measurement System PPMS (Quantum Design). In an ac susceptibility experiment, the response of the sample to an alternating ac magnetic field $H_{ac}$
\begin{equation}
H_{ac}(t) = H_{ac} e^{\imath \omega_{m} t}
\end{equation}
is measured (see Appendix~\ref{AppendixDemagnetization} for a detailed description of the theoretical framework). In the current experiments, the $T$ dependence of the ac susceptibility was always measured in field-cooled (FC) conditions, namely $H$ was always applied and modified for $T \geq 100$ K, deep inside the paramagnetic phase, and only harmonics up to $n^{*} = 3$ were recorded (see Appendix~\ref{AppendixDemagnetization}). The frequency $\nu_{m} = \omega_{m}/2\pi$ was swept in the range $10 - 10^{4}$ Hz while the amplitude of $H_{ac}$ was chosen between $0.4 - 4$ Oe. Superimposed to $H_{ac}(t)$, static polarizing magnetic fields $H$ were applied up to $25$ kOe. Similarly to the cases of dc magnetometry and LF-$\mu^{+}$SR discussed above, both $H_{ac}(t)$ and $H$ were applied parallel to the $[1 \; 1 \; 1]$ direction. Under these conditions, the demagnetization factor $D_{m}$ is the same as reported in Sect.~\ref{SectionExperimentalsDcMagnetometry}. The demagnetization-corrected expressions for the complex susceptibilities $\chi_{n}$ ($n = 1, 2, 3$) are derived in Appendix~\ref{AppendixDemagnetization}, see Eqs.~\eqref{EqCaseFirstHarmonic}, \eqref{EqCaseSecondHarmonic} and \eqref{EqCaseThirdHarmonic}. Again, only demagnetization-corrected data are considered from now on. 

\section{Results and discussion}

\subsection{dc magnetometry}\label{SectionResultsDcMagnetization}

\begin{figure}[t!] 
\vspace{6.2cm} \includegraphics{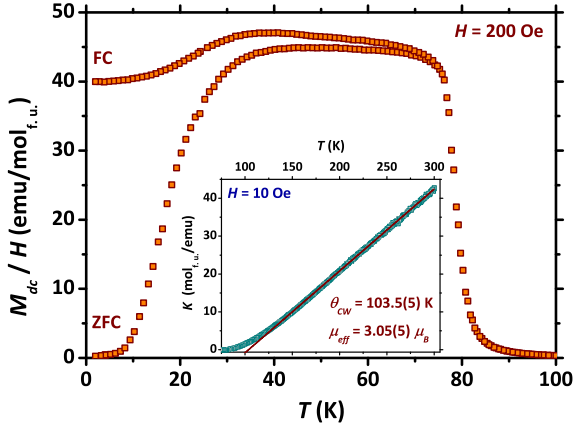}
\caption{\label{FigZFCFC200Oe}(Color online) Main panel: dc magnetic susceptibility $M_{dc}/H$ vs. $T$ at $H = 200$ Oe in both ZFC and FC conditions. A steep increase of $M_{dc}$ at around $80$ K denotes the onset of FM correlations. An irreversible opening of the two curves is observed at lower $T$. The broad decrease of $M_{dc}$ for $T \lesssim 40$ K suggests a gradual antiferromagnetic rearrangement of the Sm$^{3+}$ ions. Inset: $T$ dependence of $K \equiv \left[\chi_{dc}-\left(\chi_{dc}\right)_{0}\right]^{-1}$ at $H = 10$ Oe. The continuous line is a best fit to the experimental data according to Eq.~\eqref{EqCurieWeissFitting}. Values for the fitting parameters are reported in the figure.}
\end{figure}

The $\chi_{dc}$ vs. $T$ curves both in zero-field-cooled (ZFC) and FC conditions for $H = 200$ Oe are presented in the main panel of Fig.~\ref{FigZFCFC200Oe}. Upon cooling the sample, the steep increase in $\chi_{dc}$ denotes the onset of ferromagnetic (FM) correlations across the Mo$^{4+}$ sublattice for $T \sim 80$ K, displaying a good agreement with several previous reports on powder samples.\cite{Ali89,Tag99,Sin08} It should be immediately remarked that this \textit{quantitative} similarity can be considered as a good criterion\cite{Sil14} in order to infer a negligible degree of oxygen vacancies within the currently investigated single crystal. Oxygen off-stoichiometry is indeed a well-known problem in pyrochlore molybdates that may strongly affect the physical properties of the sample and, in particular, the $T_{C}$ value.\cite{Gar10,Cao95}
\begin{table}[b!]
\caption{Estimates of the paramagnetic moment of Sm$^{3+}$ ions under different conditions of anisotropy, namely Heisenberg, easy-axis and easy-plane (only for the case ${\bf H} \parallel [1 \; 1 \; 1]$). Heisenberg conditions are always assumed for $\mu_{Mo}$, whose chosen limiting values are relative to the range typically reported in the literature. As described in the text, $\mu_{eff}$ is the total magnetic moment per SmMoO$_{3.5}$ units.}
\label{TabSmConfigurations}%
\vspace*{0.3cm}
\bgroup
\begin{tabular}{ccc}
\hline
\hline
$\boldsymbol{\mu}_{Sm}$ configuration, $\left[\mu_{eff}\right]$ & $\mu_{Mo}$ ($\mu_{B}$) & $\mu_{Sm}$ ($\mu_{B}$)\\
\hline
\hline
\multirow{2}{*}{Heisenberg, $\left[\mu_{Mo} + \mu_{Sm}\right]$} & $\sim 2.1$ & $\sim 0.95$\\
\cline{2-3}
& $\sim 2.4$ & $\sim 0.65$\\
\hline
\multirow{2}{*}{$\parallel \langle 1 \; 1 \; 1 \rangle$, $\left[\mu_{Mo} + \left(\nicefrac{1}{2}\right) \mu_{Sm}\right]$} & $\sim 2.1$ & $\sim 1.9$\\
\cline{2-3}
& $\sim 2.4$ & $\sim 1.3$\\
\hline
\multirow{2}{*}{$\perp \langle 1 \; 1 \; 1 \rangle$, $\left[\mu_{Mo} + \left(\nicefrac{\sqrt{2}}{2}\right) \mu_{Sm}\right]$} & $\sim 2.1$ & $\sim 1.35$\\
\cline{2-3}
& $\sim 2.4$ & $\sim 0.9$\\
\hline
\hline
\end{tabular}
\egroup
\end{table}

The FM nature of correlations among Mo$^{4+}$ is confirmed by fitting the data in the paramagnetic regime $T > T_{C}$ by means of a Curie-Weiss expression
\begin{equation}\label{EqCurieWeissFitting}
\chi_{dc}(T) = \frac{C}{T-\theta_{CW}} + \left(\chi_{dc}\right)_{0}, \qquad C = \frac{N_{A} \mu^{2}}{3 k_{B}},
\end{equation}
where $\theta_{CW}$ is the Curie-Weiss temperature, $\mu = n \times \mu_{B}$ is the total magnetic moment in the paramagnetic phase expressed as $n$ times the Bohr magneton $\mu_{B}$, while $N_{A}$ and $k_{B}$ are the Avogadro number and the Boltzmann constant, respectively. The term $\left(\chi_{dc}\right)_{0}$ in Eq.~\eqref{EqCurieWeissFitting} accounts for a $T$-independent magnetic susceptibility. An illustrative fitting curve to $K \equiv \left[\chi_{dc}-\left(\chi_{dc}\right)_{0}\right]^{-1}$ data for $H = 10$ Oe is shown in the inset of Fig.~\ref{FigZFCFC200Oe}. The resulting values for the fitting parameters are $\theta_{CW} = \left(103.5 \pm 0.5\right)$ K and $\mu_{eff} = \left(\nicefrac{1}{2}\right) \mu \simeq \left(3.05\pm 0.05\right) \mu_{B}$/SmMoO$_{3.5}$. From the positive value of $\theta_{CW}$, in good agreement with previous reports,\cite{Raj92} FM correlations can be inferred for the Mo$^{4+}$ sublattice. The value for $\mu_{eff}$ is higher than a similar estimate performed for Y$_{2}$Mo$_{2}$O$_{7}$,\cite{Sil14} as expected since here the estimate involves the paramagnetic contribution of both Sm$^{3+}$ and Mo$^{4+}$ magnetic moments. A comparable value of $\mu_{eff}$ is reported for Sm$_{2}$Mo$_{2}$O$_{7}$ in an overview of pyrochlore molybdates presented in Ref.~\onlinecite{Ran83}, while a slightly lower value was reported elsewhere.\cite{Raj92} A theoretical value $\mu_{Mo} \simeq 2.8 \; \mu_{B}$ for the paramagnetic moment of Mo$^{4+}$ has been reported, while the measured values are typically in the range $2.1 - 2.4 \; \mu_{B}$.\cite{Sil14,Raj92,Gar99,Ehl10} Thus, different estimates for $\mu_{Sm}$ can be deduced for the paramagnetic moment of Sm$^{3+}$ by assuming different anisotropic properties. In this respect, as already stressed in the introduction, it should be remarked that this topic is still highly controversial due to the absence of reliable experimental data. Local easy-axis magnetic anisotropy along $\langle 1 \; 1 \; 1 \rangle$ directions for the Sm$^{3+}$ moments is claimed in Ref.~\onlinecite{Sin08}. At the same time, from theoretical arguments about crystal field potential on the basis of the Stevens' operator equivalents and, in particular, on the sign of the $B_{2}^{0}$ coefficient, an easy-plane configuration for Sm$^{3+}$ magnetic moments could be envisaged more likely.\cite{Gar10} In Tab.~\ref{TabSmConfigurations}, we report estimates for the value of the paramagnetic $\mu_{Sm}$ under different conditions of anisotropy (and in the case of interest ${\bf H} \parallel [1 \; 1 \; 1]$). Heisenberg conditions are always assumed for $\mu_{Mo}$, whose values are chosen from the typical range reported in the literature.

\begin{figure}[t!] 
\vspace{6.2cm} \includegraphics{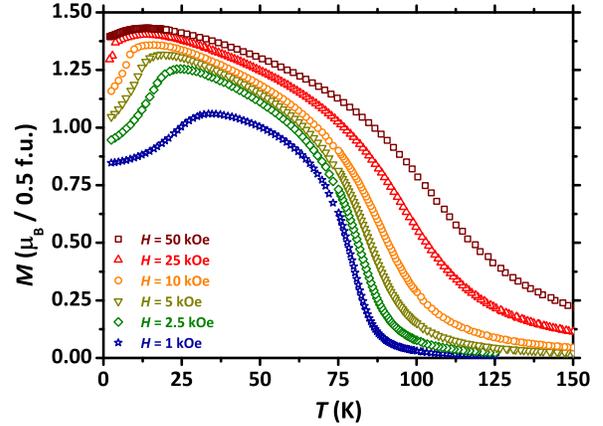}
\caption{\label{FigFigMvsT}(Color online) $M_{dc}$ vs $T$ curves (FC) at different nominal values of $H$.}
\end{figure}
At lower $T$ values ($T \lesssim 40$ K), a broad decrease in $\chi_{dc}$ suggests a tendency of Sm$^{3+}$ magnetic moments to rearrange antiferromagnetically (AFM) with respect to the molecular field generated by the Mo$^{4+}$ sublattice.\cite{Ali89,Sin08} $M_{dc}$ vs. $T$ curves at different values of $H$ are reported in Fig.~\ref{FigFigMvsT}. Overall, the agreement with previous data reported in Ref.~\onlinecite{Sin08} is good, even if we do not observe the dramatic drop of the magnetization at low $T$ and for $H = 1$ kOe. Instead, a weaker but steady decrease of $M_{dc}$ is measured at all the considered $H$ values. The amplitude of the magnetization drop is reduced by increasing $H$, which may be explained by considering the competing effect of increasing $H$ and of Sm-Mo mutual AFM correlations. With increasing $H$, up to $H = 50$ kOe, the curves clearly saturate towards an ordered value corresponding to $\sim 1.4 \; \mu_{B}$/SmMoO$_{3.5}$.\cite{Sin08}It should be pointed out that the reduction of the paramagnetic moment within the ordered FM phases is a well-known effect for itinerant systems\cite{Tak86,Oht09} and it may be playing a role also in the current case for Mo$^{4+}$.

\begin{figure}[t!] 
\vspace{6.2cm} \includegraphics{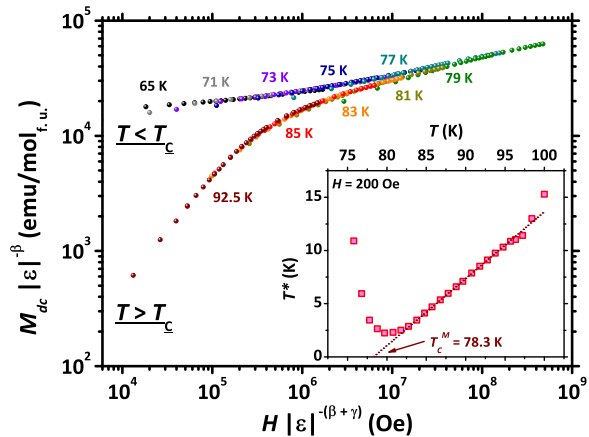}
\caption{\label{FigScalingMdc}(Color online) Inset: function $T^{*}(T)$, described in the text [see Eq.~\eqref{EqKouvelFisherTc}], as obtained from the experimental data of $\chi_{dc}$ reported in Fig.~\ref{FigZFCFC200Oe} (paramagnetic phase). The dotted line is a best fitting function according to Eq.~\eqref{EqKouvelFisherTc}, allowing the estimates $T_{C}^{M} = \left(78.3 \pm 0.1\right)$ K and $\gamma = 1.59 \pm 0.03$ (see text). Main panel: scaling behaviour of isothermal $M_{dc}$ vs. $H$ curves at different $T$ values around $T_{C}^{M}$, as described in the text [see Eq.~\eqref{EqScalingFormalism}]. By considering the estimates performed in the paramagnetic phase as fixed parameters [see Eq.~\eqref{EqKouvelFisherTc} and inset], a value $0.345 \pm 0.02$ is estimated for the critical exponent $\beta$.}
\end{figure}
The data presented in Fig.~\ref{FigZFCFC200Oe} are analyzed according to the Kouvel-Fisher formalism.\cite{Kou64,Das93,Nai03,Bha11,Tsu13,Tat14} This framework allows to obtain a precise estimate of the critical temperature $T_{C}^{M}$ for the transition of the Mo$^{4+}$ sublattice, the superscript $M$ referring to the dc magnetometry technique (see the inset of Fig.~\ref{FigScalingMdc}). In particular, one should expect a linear dependence on $T$ for the function
\begin{equation}\label{EqKouvelFisherTc}
T^{*}(T) = \frac{d\left[\ln\left(\chi_{dc}^{-1}\right)\right]}{dT} \propto \frac{T}{\gamma}
\end{equation}
in the paramagnetic phase.\cite{Kou64,Das93,Nai03,Bha11,Tsu13,Tat14} Here, the slope is determined by the critical exponent $\gamma$ characteristic of the transition, while the relation $T^{*}(T_{C}^{M}) = 0$ holds. The expected linear trend is observed in our data for Sm$_{2}$Mo$_{2}$O$_{7}$ (see the inset of Fig.~\ref{FigScalingMdc}), allowing the estimates $T_{C}^{M} = \left(78.3 \pm 0.1\right)$ K and $\gamma = 1.59 \pm 0.03$. It should be stressed that typically $\gamma \simeq 1.2 - 1.4$ for the three-dimensional universality classes,\cite{Tat14,Kau85} while values compatible with our observation are usually reported for  AmFA.\cite{Kau85} The $T_{C}^{M}$ value is in very good agreement with other reports in the literature from magnetic techniques, while a slightly higher value $T_{C} \simeq 86$ K was reported by means of resistivity on sintered powder samples.\cite{Gre87}

The critical behaviour of the Mo$^{4+}$ ordered phase in the proximity of $T_{C}^{M}$ can be investigated in more detail by a closer examination of the isothermal $M_{dc}$ vs. $H$ curves. In particular, according to the scaling formalism,\cite{Das93,Bha11,Tsu13,Tat14,Kau85,Man83} such curves should collapse onto two different well-defined branches described by the functions $f_{+}$ (for $T > T_{C}^{M}$) and $f_{-}$ (for $T < T_{C}^{M}$) defined by the relation
\begin{equation}\label{EqScalingFormalism}
m = f_{\pm}\left(h\right) \qquad \textrm{where} \qquad
\begin{cases}
m \equiv M_{dc} \left|\varepsilon\right|^{-\beta}\\
h \equiv H \left|\varepsilon\right|^{-\left(\beta+\gamma\right)}
\end{cases}
.
\end{equation}
Here, $\beta$ and $\gamma$ are the critical exponents characteristic of the transition whereas $\varepsilon = \left(T-T_{C}^{M}\right)/T_{C}^{M}$ is the reduced temperature. As it is shown in the main panel of Fig.~\ref{FigScalingMdc}, this scaling behaviour can be well-reproduced in the current case of Sm$_{2}$Mo$_{2}$O$_{7}$ by keeping the two values obtained above for $T_{C}^{M}$ and $\gamma$ constant and yielding $\beta = 0.345 \pm 0.02$, a value which is consistent with the three-dimensional character of the magnetic correlations.\cite{Tat14} In turn, by the Widom relation $\delta = 1 + \left(\gamma/\beta\right)$, one has $\delta = 5.61 \pm 0.45$ for the third critical exponent of the magnetic phase transition.\cite{Sta71} Also in this case, $\delta$ is higher than what is typically reported for three-dimensional ferromagnets, namely $\sim 4.8$,\cite{Tat14} hinting at a closer analogy between currently investigated Sm$_{2}$Mo$_{2}$O$_{7}$ and AmFA.\cite{Kau85}

It should be pointed out that some discrepancies from the values reported above are obtained if other criteria are considered (isotherm $M_{dc}$ vs. $H$ curve at $T_{C}^{M}$, modified Arrott plots).\cite{Arr57,Arr67,Bha11,Tsu13,Tat14} Since these different methods involve only $M_{dc}$ vs. $H$ curves, it is likely that the origin of such discrepancies stems from the uncertainty in the demagnetization factor $D$. However, as it is shown in the next sections, our estimates reported above are confirmed by means of different independent experimental techniques.

\subsection{Muon spin spectroscopy}\label{SectionResultsMuSR}

\begin{figure}[t!] 
\vspace{6.2cm} \includegraphics{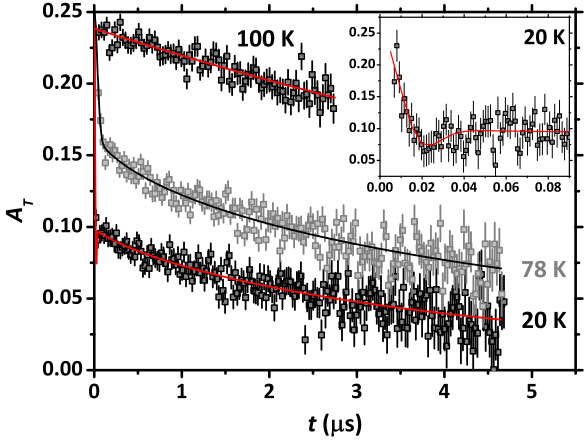}
\caption{\label{FigMuonDepolarization}(Color online) ZF-$\mu^{+}$SR $t$-depolarization for Sm$_{2}$Mo$_{2}$O$_{7}$ at selected $T$ values. The short $t$ behaviour of the curve for $T = 20$ K is enlarged in the inset, evidencing the strongly-damped oscillation.}
\end{figure}

Representative ZF-$\mu^{+}$SR $t$-depolarizations for Sm$_{2}$Mo$_{2}$O$_{7}$ are shown in Fig.~\ref{FigMuonDepolarization}. The development of magnetic correlations of electronic origin within a bulk fraction of the sample is clearly observed for $T \lesssim 80$ K, in agreement with measurements of dc magnetometry reported in Sect.~\ref{SectionResultsDcMagnetization}. As noticed at very small $t$ values (see the inset of Fig.~\ref{FigMuonDepolarization}), a clear dip appears for all $T$ values below $\sim 80$ K, not detected in a previous work on Sm$_{2}$Mo$_{2}$O$_{7}$\cite{Jo05} (possibly due to different experimental conditions and worse $t$ resolution). In the case of a conventional long-range FM phase, one would typically detect long-lived coherent oscillations for the ZF-$\mu^{+}$SR $t$-depolarization,\cite{Pra13a,Pra13b} as actually reported for several metallic pyrochlores.\cite{Dis12a,Dis12b,Guo13,Dis14} Accordingly, this short-$t$ feature can be interpreted as an overdamped oscillation, revealing a wide distribution of local fields $B_{\mu}$ at the $\mu^{+}$ site, namely, a severe degree of magnetic disorder. Indeed, the shape of the $t$-depolarization is highly reminiscent of the well-known Kubo-Toyabe functions typically observed in spin glasses\cite{Kub81,Uem85,Gin97a,Yao11} and in the isostructural compounds (Tb$_{1-x}$La$_{x}$)$_{2}$Mo$_{2}$O$_{7}$, Gd$_{2}$Mo$_{2}$O$_{7}$ and  Y$_{2}$Mo$_{2}$O$_{7}$.\cite{Ape06,Mir06,Dun96} However, the accepted phase diagram of pyrochlore molybdates confines the spin-glass behaviour to the insulating region\cite{Igu09,Gar10,Shi13} or, at least, close enough to the MIT boundary.\cite{Kat00} Clearly, our current data do not match within the currently accepted framework of a sharp MIT for $R_{2}$Mo$_{2}$O$_{7}$.

\begin{figure}[t!] 
\vspace{11.2cm} \includegraphics{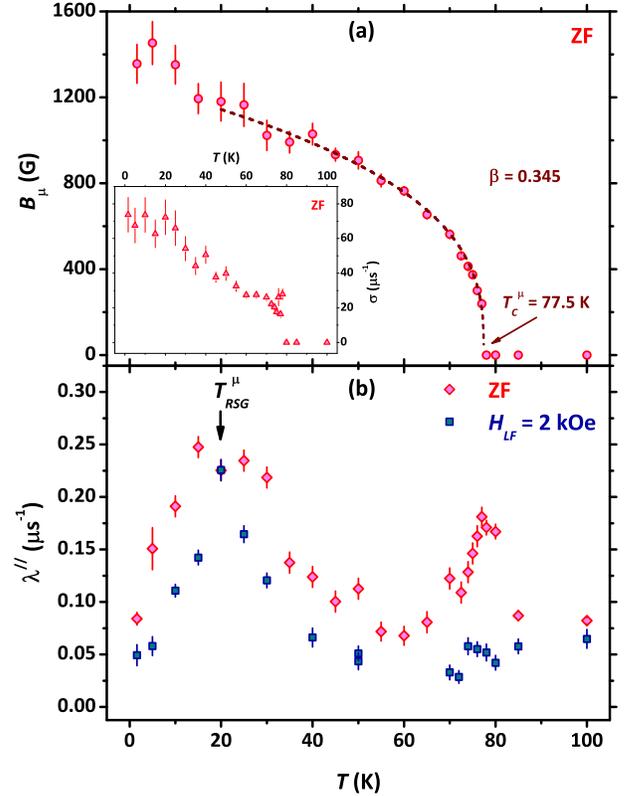}
\caption{\label{FigInternalFieldAndRelaxation}(Color online) Panel (a): $T$ dependence of the internal magnetic field $B_{\mu}$. The dashed line is a best fit to our experimental data according to Eq.~\eqref{EqPowerLawInternalField}. The value of $\beta$ has been kept fixed to the estimate obtained by the scaling analysis of $M_{dc}$ (see Fig.~\ref{FigScalingMdc}). Inset of panel (a): $T$ dependence of the transverse relaxation $\sigma$ for our ZF data. Panel (b): $T$ dependence of the longitudinal relaxation $\lambda^{\parallel}$ under different conditions for the external longitudinal field. The characteristic temperature $T_{_{RSG}}^{\mu} \simeq 20$ K can be defined as the maximum of the broad peak at low $T$.}
\end{figure}
Fitting our experimental curves by means of either purely-Gaussian or purely-Lorentzian Kubo-Toyabe functions does not yield satisfactory results across the whole investigated $T$ range, similarly to the cases of (Tb$_{1-x}$La$_{x}$)$_{2}$Mo$_{2}$O$_{7}$ and Y$_{2}$Mo$_{2}$O$_{7}$.\cite{Ape06,Dun96} Accordingly, a more phenomenological approach is employed by referring to the conventional formula for ordered magnetic materials\cite{Ape06,Mir06,Pra13b,Car13}
\begin{eqnarray}\label{EqGeneralFittingZFSample}
G_{T}^{s}(t) & = & e^{-\left(\frac{\sigma_{N}^{2} t^{2}}{2} + \lambda_{e} t\right)} \left\{ \left[1 - V_{m}(T)\right] +\right. {}\\ & + & \left.\left[a^{\perp}(T) \cos\left(\gamma_{\mu} B_{\mu} t\right) e^{-\frac{\left(\sigma t\right)^{2}}{2}} + a^{\parallel}(T) e^{-\left(\lambda^{\parallel}t\right)^{\alpha}}\right]\right\}\nonumber
\end{eqnarray}
[see Eq.~\eqref{EqGeneralFittingZF}], where the parameter $V_{m}(T)$ quantifies the fraction of $\mu^{+}$ probing a static local magnetic field of electronic origin. In the paramagnetic limit, $V_{m}(T) = 0$, no static field of electronic origin contributes to the depolarization and only the weak field distribution induced by the nuclear magnetic moments leads to a slow Gaussian depolarization governed by the rate $\sigma_{N}$. Superimposed to the weak nuclear contribution, the exponential term $\lambda_{e}$ accounts for all the possible sources of $T$-dependent dynamical relaxation (e.~g., diluted magnetic impurities, $\mu^{+}$ hopping/diffusion).\cite{Yao11} Below $T_{C}$, the superscript $\perp$ ($\parallel$) refers to $\mu^{+}$ experiencing a local static magnetic field in a perpendicular (parallel) direction with respect to the initial $\mu^{+}$ spin polarization, while the parameters $a$ quantify their relative weights within the overall signal, with
\begin{equation}
a^{\perp}(T) + a^{\parallel}(T) = V_{m}(T).
\end{equation}
The $a^{\perp}$ ($a^{\parallel}$) fraction is referred to as transverse (longitudinal) in the following. In the presence of a long-range magnetic order, a coherent precession of $\mu^{+}$ around the local magnetic field $B_{\mu}$ can be discerned in the $a^{\perp}$ fraction and described by the oscillating $\cos$-like function (where $\gamma_{\mu} = 2 \pi \times 135.54$ MHz/T is the gyromagnetic ratio for $\mu^{+}$). The relative Gaussian (over)damping term governed by the rate $\sigma$ reflects a distribution of local magnetic field values at the $\mu^{+}$ site. On the other hand, the $a^{\parallel}$ component is typically damped by the exponentially decaying function governed by $\lambda^{\parallel}$ and it probes spin-lattice relaxation processes. Finally, the stretching parameter $\alpha$ accounts for a distribution of relaxation rates, typical of disordered glassy magnets.\cite{Cam94} One single possible crystallographic position is assumed for $\mu^{+}$.

Results of the fitting to raw experimental data are shown for different selected $T$ values in Fig.~\ref{FigMuonDepolarization} and the $T$ dependence of the most important fitting parameters is reported in Fig.~\ref{FigInternalFieldAndRelaxation}. The internal magnetic field $B_{\mu}$ [see Fig.~\ref{FigInternalFieldAndRelaxation}, panel (a)] displays a very good agreement with the scaling analysis of our dc magnetization data since it can be fit to a power-law expression
\begin{equation}\label{EqPowerLawInternalField}
B_{\mu}(T) = B_{\mu}(0) \left(1 - \frac{T}{T_{C}^{\mu}}\right)^{\beta}
\end{equation}
using $\beta = 0.345$ as the critical exponent for the ZF internal magnetization.\cite{Sta71} The fit leads to a slightly lower value for $T_{C}^{\mu} = 77.5 \pm 0.2$ K if compared to the estimate from the Kouvel-Fisher analysis of our $M_{dc}$ data. It is interesting to stress that the $\mu^{+}$ spin depolarization does not show qualitative differences at any $T$ value below $\sim 75$ K and, in particular, for $T \lesssim 40$ K, besides a small departure of the internal field from the power-law trend described by Eq.~\eqref{EqPowerLawInternalField}. This is in agreement with the scenario of a gradual reorientation of Sm$^{3+}$ magnetic moments rather than a phase transition for that sublattice, as suggested by the anomalies in $M_{dc}$ vs. $T$ curves (see Fig.~\ref{FigFigMvsT} and Ref.~\onlinecite{Sin08}).

Results for the transverse relaxation $\sigma$ are reported in the inset of Fig.~\ref{FigInternalFieldAndRelaxation}, panel (a), while the longitudinal relaxation $\lambda^{\parallel}(T)$ is reported for both ZF and LF ($H_{LF} = 2$ kOe) data in Fig.~\ref{FigInternalFieldAndRelaxation}, panel (b). The ZF data clearly evidence two different peaks for these latter data, in qualitative agreement with previous reports on Sm$_{2}$Mo$_{2}$O$_{7}$\cite{Jo05} and, also, on (Tb$_{1-x}$La$_{x}$)$_{2}$Mo$_{2}$O$_{7}$.\cite{Ape06} The sharp one around $T \sim 80$ K is due to the critical dynamics associated with the transition of the Mo$^{4+}$ sublattice and, remarkably, it is completely quenched by the application of $H_{LF} = 2$ kOe. The much broader peak at lower $T$ values reveals interesting insights in the disordered magnetic state. After comparison with $M_{dc}$ data discussed in Sect.~\ref{SectionResultsDcMagnetization}, it may be tempting to assign such dynamical features to the re-orientation of Sm$^{3+}$ magnetic moments. However, the strong $H$-dependence of anomalies in $M_{dc}$ vs. $T$ curves should be pointed out, especially in the low-$H$ limit. This is opposite to what displayed by the dynamical peak in Fig.~\ref{FigInternalFieldAndRelaxation}(b), whose typical temperature at the maximum does not shift with $H$. Remarkably, the two-peaks structure is very much similar to what reported in the case of Fe$_{1-x}$Mn$_{x}$ AmFA, instead.\cite{Gin97a,Mir97} This points towards a freezing of the transverse $XY$ spin components of Mo$^{4+}$ in the $T \sim 25$ K region for Sm$_{2}$Mo$_{2}$O$_{7}$, an effect otherwise known as re-entrant spin glass (RSG).\cite{Gin97a} This is an invaluable information brought by $\mu^{+}$SR, since $XY$ freezing is expected not to contribute to macroscopic magnetization.\cite{Gin97a} Accordingly, the $T$ value corresponding to the low $T$ maximum in $\lambda^{\parallel}$ is defined as $T_{_{RSG}}^{\mu}$ and it takes a value $T_{_{RSG}}^{\mu} = \left(20 \pm 2\right)$ K. More insights into the RSG phase at low $T$ values, as obtained by means of magnetic ac susceptibility, are presented in the next section.

\subsection{Magnetic ac susceptibility}

\begin{figure}[t!]
\vspace{6.2cm} \includegraphics{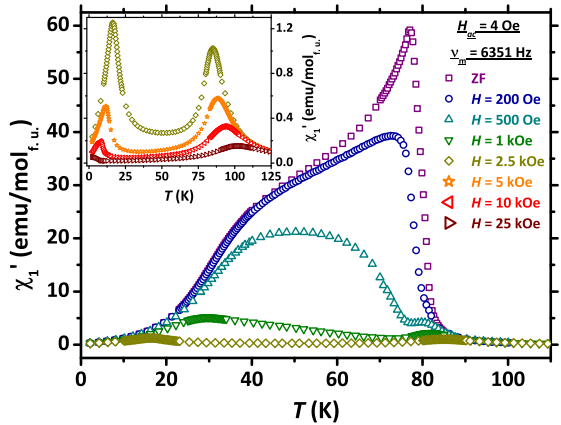}
\caption{\label{FigAt6351HzFirstHarmonic}(Color online) $\chi_{1}^{\prime}$ vs. $T$ curves at different $H$ values. $H$ values reported in the legend are nominal values not corrected by demagnetization effects. Notice the different $y$-axis scale between main panel and inset. Measurements relative to the fixed values of $\nu_{m} = 6351$ Hz and $H_{ac} = 4$ Oe are reported.}
\end{figure}

Experimental results for the real component of the first-harmonic ac magnetic susceptibility $\chi_{1}$ are shown in Fig.~\ref{FigAt6351HzFirstHarmonic}. In ZF, $\chi_{1}^{\prime}$ shows a sharp peak with a maximum at $T_{C}^{\chi} = 77.0 \pm 0.1$ K followed by a broad decrease for decreasing $T$. The overall $\chi_{1}^{\prime}$ contribution is suppressed upon increasing the value of the polarizing magnetic field $H$ (see Fig.~\ref{FigAt6351HzFirstHarmonic}). This behaviour is very much similar to what is typically observed for AmFA,\cite{Maa76,Maa78,Ho81a,Ho81b,Ber95} thus confirming again the similarities already stressed in the previous sections concerning $M_{dc}$ and $\mu^{+}$SR data. 

This strong similarity with AmFA can be put on a more quantitative basis by a closer investigation of the main contributions to $\chi_{1}^{\prime}$. As it is shown in the inset of Fig.~\ref{FigAt6351HzFirstHarmonic}, only two distinct peaks are left in $\chi_{1}^{\prime}$ vs. $T$ for $H \gtrsim 1$ kOe indeed. These peaks behave differently upon increasing $H$. In particular, the low-$T$ peak shifts to even lower $T$ values with increasing $H$ while the contrary is true for the high-$T$ peak (see also the enlargement of data later in the main panels of Figs.~\ref{FigAt6351HzFirstHarmonicZoomLTAndScaling} and \ref{FigAt6351HzFirstHarmonicFirZoomHTAndScaling}, respectively). The origin of these anomalies is discussed in detail in the next subsections.

\subsubsection{High-$T$ critical peak}

\begin{figure}[t!]
\vspace{6.2cm} \includegraphics{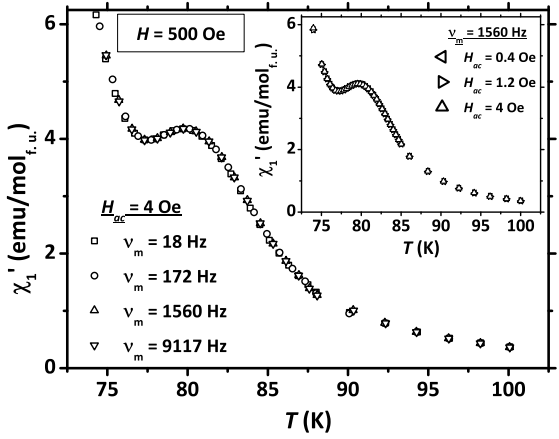}
\caption{\label{FigIndepNuHacHighTPeakFirstHarmonic} Enlargement of $\chi_{1}^{\prime}$ vs. $T$ data (see Fig.~\ref{FigAt6351HzFirstHarmonic}) around the critical peak at the fixed value $H = 500$ Oe. Curves are reported at the fixed value $H_{ac} = 4$ Oe and different $\nu_{m}$ (main panel) and at the fixed value $\nu_{m} = 1560$ Hz and different $H_{ac}$ (inset). No dependence is inferred on both $\nu_{m}$ and $H_{ac}$.}
\end{figure}

The high-$T$ peak is associated with the critical divergence of magnetic fluctuations in the proximity of the transition of the Mo$^{4+}$ sublattice (hence the name ``critical peak'').\cite{Maa78,Ho81a,Ho81b,Ber95} For each $H$ value, the data have been measured for different values of both $H_{ac}$ and $\nu_{m}$. As it is shown in Fig.~\ref{FigIndepNuHacHighTPeakFirstHarmonic}, no dependence at all on both parameters was detected for the experimental points, as expected for the critical peak.\cite{Maa78,Ho81a,Ho81b,Ber95}

\begin{figure}[t!]
\vspace{6.2cm} \includegraphics{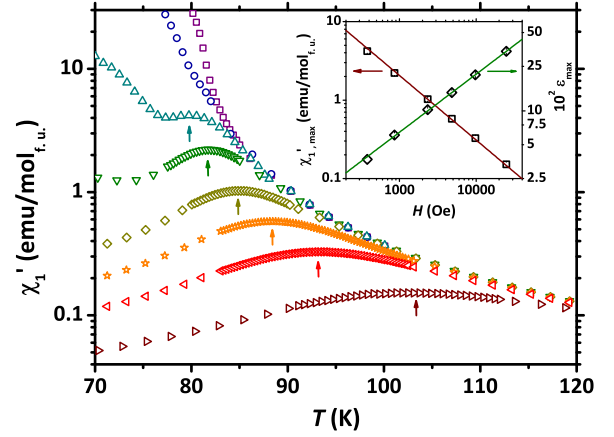}
\caption{\label{FigAt6351HzFirstHarmonicFirZoomHTAndScaling}(Color online) Main panel: enlargement in the high-$T$ region of $\chi_{1}^{\prime}$ vs. $T$ data already presented in Fig.~\ref{FigAt6351HzFirstHarmonic} evidencing the $H$ dependence of the critical peak. Same symbols as in Fig.~\ref{FigAt6351HzFirstHarmonic} are used. Inset: $H$ dependence of the peak value of $\chi_{1}^{\prime}$ and of the corresponding reduced temperature values $\varepsilon_{max}$ (empty squares and empty diamonds, respectively. See the arrows in the main panel). The continuous lines reproduce the power-law trends reported in Eq.~\eqref{EqScalingAcSusceptibility} where the critical exponents are kept fixed to the values estimated from the scaling analysis of $M_{dc}$ data. Values of $H$ on the $x$ axis are reported after correction for the demagnetization factor [see Eq.~\eqref{EqDemagFactMvsH}].}
\end{figure}

Scaling analysis shows that the critical exponent $\delta$ governs the rate of suppression of $\chi_{1,max}^{\prime}$ (namely, the maximum value of $\chi_{1}^{\prime}$ within the critical peak) as a function of $H$, the latter being corrected in order to take demagnetization effects into account. At the same time, the $H$ dependence of the temperature value $T_{max}$ corresponding to the maximum in $\chi_{1}^{\prime}$ also scales with the intrinsic value of the magnetic field at a rate governed by $\beta$ and $\gamma$. In particular, one has\cite{Maa78,Ho81a,Ho81b,Ber95}
\begin{eqnarray}\label{EqScalingAcSusceptibility}
\chi_{1,max}^{\prime}(H) & \propto & H^{-1+\left(1/\delta\right)},{}\nonumber\\ \frac{T_{max}(H)-T_{C}^{\chi}}{T_{C}^{\chi}} = \varepsilon_{max} & \propto & H^{\left(\beta+\gamma\right)^{-1}}.
\end{eqnarray}
Data for the $H$ dependence of both $\chi_{1,max}^{\prime}$ and $\varepsilon_{max}$ are reported in the inset of Fig.~\ref{FigAt6351HzFirstHarmonicFirZoomHTAndScaling}, together with two curves according to Eq.~\eqref{EqScalingAcSusceptibility} with the critical exponent values fixed to those estimated in Sect.~\ref{SectionResultsDcMagnetization} (see the continuous lines). The agreement with our experimental data is remarkable, again confirming the correctness of the overall framework for the estimate of the critical exponents of Sm$_{2}$Mo$_{2}$O$_{7}$. A summarizing set of data for the values of $T_{C}$ estimated from the different experimental techniques and the critical exponents estimated from dc magnetometry is reported in Tab.~\ref{TabParametersFMtransition}.
\begin{table}[b!]
\caption{Estimates of $T_{C}$ and of critical exponents for the transition of the Mo$^{4+}$ sublattice as obtained from the different experimental techniques ($M$: dc magnetometry, $\mu$: $\mu^{+}$SR, $\chi$: ac susceptibility).}
\label{TabParametersFMtransition}%
\vspace*{0.3cm}
\bgroup
\begin{tabular}{ccc}
\hline
\hline
&\phantom{a}\phantom{a}  $78.3 \pm 0.1$ & ($M$)\\
$T_{C}$ (K) &\phantom{a}\phantom{a}  $77.5 \pm 0.2$ & ($\mu$)\\
&\phantom{a}\phantom{a}  $77.0 \pm 0.1$ & ($\chi$)\\
\hline
$\gamma$ &\phantom{a}\phantom{a}   $1.59 \pm 0.03$ & ($M$)\\
\hline
$\beta$ &\phantom{a}\phantom{a}   $0.345 \pm 0.02$ & ($M$)\\
\hline
$\delta = 1 + \left(\gamma/\beta\right)$ &\phantom{a}\phantom{a}   $5.61 \pm 0.45$ & ($M$)\\
\hline
\hline
\end{tabular}
\egroup
\end{table}

\begin{figure*}[htbp]
\vspace{6.2cm} \includegraphics{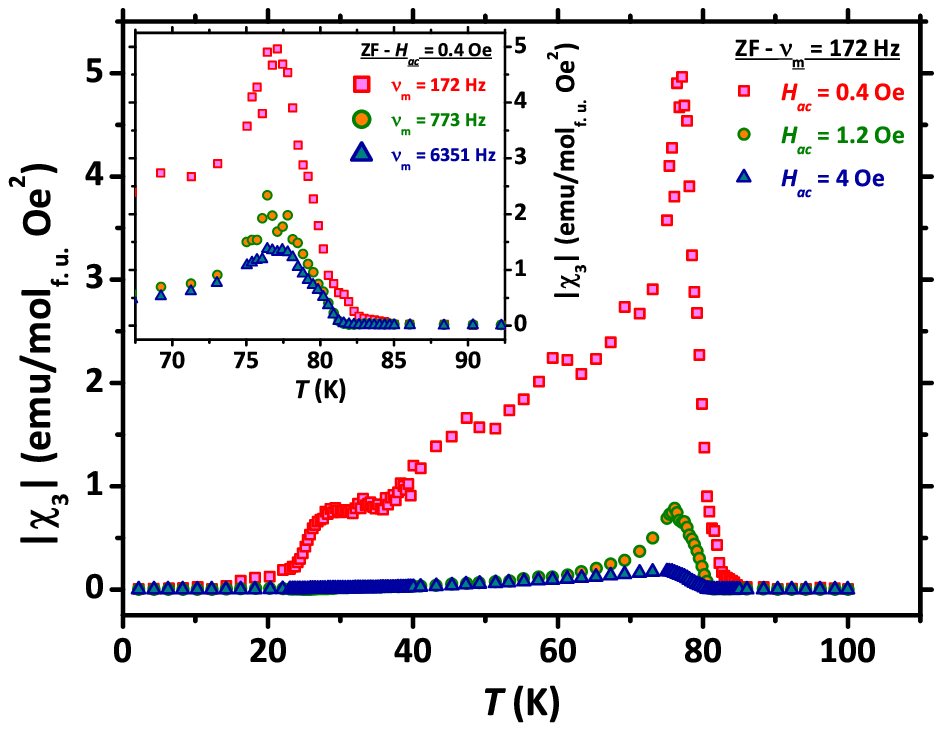} \hspace*{1.2cm} \includegraphics{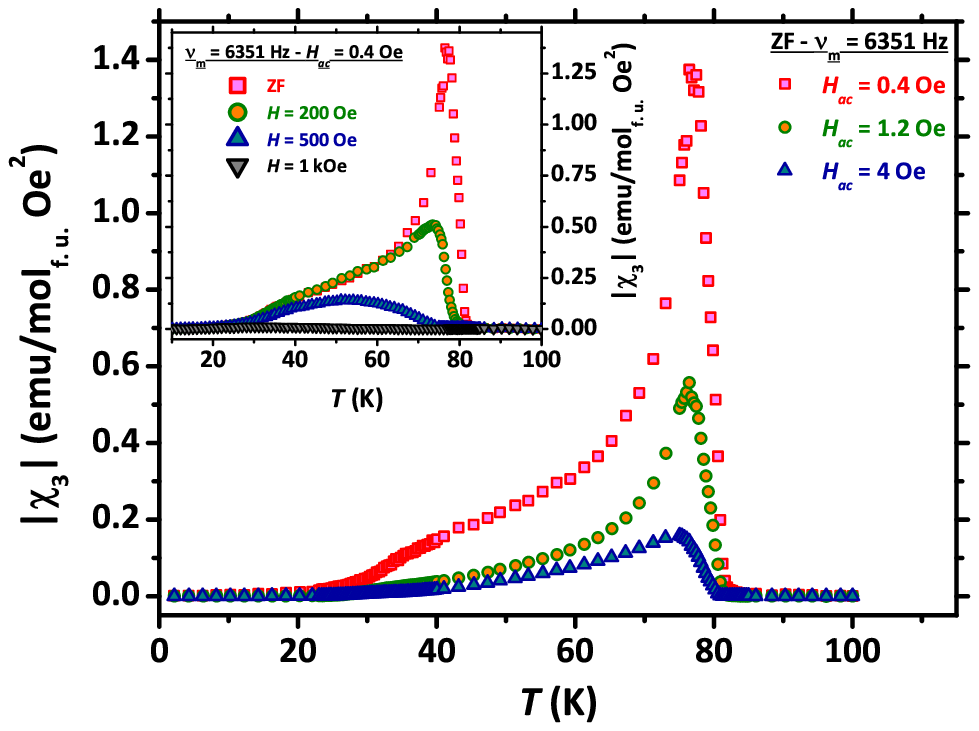} 
\caption{\label{FigsThirdHarmonic} (Color online) $T$ dependence of $\left|\chi_{3}\right|$ [see Eq.~\eqref{EqDefinitionModulusThirdHarmonic}] for different experimental conditions (see legends in the different panels). The label ZF refers to the case of zero polarizing static magnetic field, namely $H = 0$.}
\end{figure*}
Further information about the Mo$^{4+}$ glassy magnetic phase can be obtained from the analysis of the non-linear magnetic susceptibility,\cite{Gin96,Gin97} as accessed by the analysis of higher-order harmonics ac susceptibilities\cite{Miy79,Chi79,Bin86,Baj97} (see details in Appendix~\ref{AppendixDemagnetization}). Data for $\chi_{3}$, in particular, are reported in Fig.~\ref{FigsThirdHarmonic} where the absolute value
\begin{equation}\label{EqDefinitionModulusThirdHarmonic}
\left|\chi_{3}\right| = \sqrt{\left(\chi_{3}^{\prime}\right)^{2} + \left(\chi_{3}^{\prime\prime}\right)^{2}}
\end{equation}
is plotted as a function of $T$ and for different values of the experimental parameters ($\nu_{m}$, $H_{ac}$, $H$). Remarkably, a distinct peak in $\left|\chi_{3}\right|$ can be clearly evidenced around the transition temperature of Mo$^{4+}$. Similarly to the case of Li$_{x}$Ni$_{1-x}$O spin glasses,\cite{Baj97} the height of the peak is strongly dependent on external parameters, displaying a strong divergence with decreasing the frequency $\nu_{m}$ and the values of the magnetic fields $H_{ac}$ and $H$ (see Figs.~\ref{FigsThirdHarmonic}).\cite{Nai03} Within these conditions, the investigated magnetic system can be unambiguously characterized as glassy.\cite{Nai03,Baj97,Miy79,Chi79} Indeed, the divergence of the non-linear susceptibility was employed as a definitive proof of glassy magnetism also in the case of Y$_{2}$Mo$_{2}$O$_{7}$.\cite{Gin96,Gin97} Different results are expected in the case of magnetic blocking.\cite{Nai04} In this latter case, non-zero values for $\left|\chi_{3}\right|$ are still detected but their dependence on external parameters is not critically diverging.\cite{Nai04}

\subsubsection{Low-$T$ peak}

\begin{figure}[b!]
\vspace{6.2cm} \includegraphics{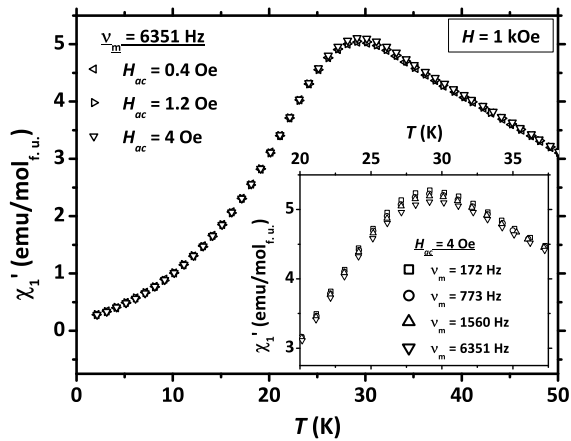} \caption{\label{FigIndepNuHacLowTPeakFirstHarmonic} Enlargement of $\chi_{1}^{\prime}$ vs. $T$ data around the low-$T$ peak at the fixed value $H = 1000$ Oe (see Fig.~\ref{FigAt6351HzFirstHarmonic}). Curves are reported at the fixed value $H_{ac} = 4$ Oe and different $\nu_{m}$ (inset) and at the fixed value $\nu_{m} = 6351$ Hz and different $H_{ac}$ (main panel). No dependence is inferred on both $\nu_{m}$ and $H_{ac}$.}
\end{figure}
\begin{figure}[t!]
\vspace{6.2cm} \includegraphics{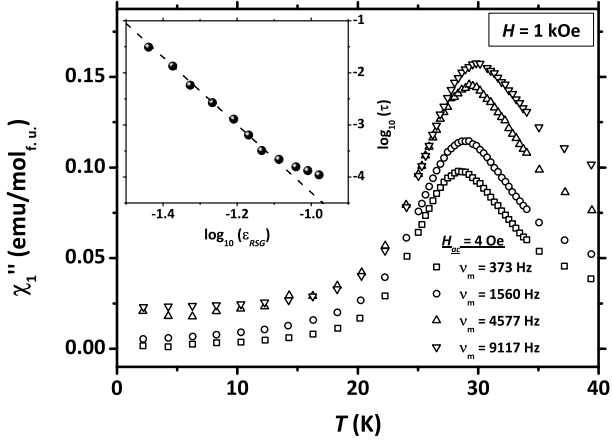}
\caption{\label{FigFreqDepLowTPeakPossibleDomains} Main panel: $T$ dependence of the imaginary component of the first-harmonic ac susceptibility $\chi_{1}^{\prime\prime}$ at fixed values $H = 1$ kOe and $H_{ac} = 4$ Oe and at different $\nu_{m}$ values. A residual dynamics can be deduced from the $T$ shift of the maximum $T_{p}(\nu_{m})$. Inset: frequency dependence of $T_{p}$. The dashed line is a best fit according to Eq.~\eqref{EqPowerLawFreqDep}. The deviation at high frequency values should be associated to non-linear effects.}
\end{figure}

In Fig.~\ref{FigIndepNuHacLowTPeakFirstHarmonic}, an enlargement is shown in the low-$T$ region for $\chi_{1}^{\prime}$ vs. $T$ data already presented in Fig.~\ref{FigAt6351HzFirstHarmonic}. In particular, the selected $T$ window focusses on the low-$T$ peak appearing for $H \gtrsim 1$ kOe. It is clearly shown that varying $H_{ac}$ does not affect $\chi_{1}^{\prime}$. A weak dependence of the amplitude of $\chi_{1}^{\prime}$ is obtained upon sweeping $\nu_{m}$ over almost three orders of magnitude. However, no dependence of the $T$ position of the peak can be extracted within the experimental error. Still, one has to notice that a weak frequency dependence is detected in the imaginary component of the first-harmonic susceptibility $\chi_{1}^{\prime\prime}$ instead (see Fig.~\ref{FigFreqDepLowTPeakPossibleDomains}). This scenario is again very similar to what is discussed in Ref.~\onlinecite{Jon96} for a RSG phase in Fe$_{1-x}$Ni$_{x}$ AmFA.

To further proof this analogy, the frequency dependence of the peak position $T_{p}(\nu_{m})$ can be examined more closely. By defining the correlation time $\tau = 1/\nu_{m}$ and the reduced temperature $\varepsilon_{_{RSG}}(\nu_{m}) = \left[T_{_{P}}(\nu_{m})-T_{_{RSG}}^{\chi}\right]/T_{_{RSG}}^{\chi}$ for the re-entrant glassy phase, one should expect a power-law trend for a glassy transition, namely
\begin{equation}\label{EqPowerLawFreqDep}
\tau = \tau_{0} \times \left[\varepsilon_{_{RSG}}(\nu_{m})\right]^{-z\nu}.
\end{equation}
The inset of Fig.~\ref{FigFreqDepLowTPeakPossibleDomains} shows that this is the case indeed. A best fit to experimental data according to Eq.~\eqref{EqPowerLawFreqDep} gives a satisfactory agreement across the (almost) whole investigated range. The resulting fitting parameters are $T_{_{RSG}}^{\chi} = 27.1$ K, $z\nu = 6.5$ and $\tau_{0} \simeq 10^{-11}$ s. Some comment is required about these values. In particular, a power-law trend can be reproduced upon modifying $T_{_{RSG}}^{\chi}$ over a very wide range (in the order of few K), with a corresponding sizeable change in the critical exponents and $\tau_{0}$ as well. Accordingly, the actual $T_{_{RSG}}^{\chi}$ value will not be considered any longer and $T_{_{RSG}}^{\mu}$ will be assumed as reliable for the final phase diagram (presented later in Fig.~\ref{FigPhaseDiagram}). At the same time, the small deviation of experimental points from the expected trend, as observed in the inset of Fig.~\ref{FigFreqDepLowTPeakPossibleDomains}, should be attributed to non-linear effects due to a too high $H_{ac}$ value. Indeed, from a study of the $H_{ac}$ dependence of $\chi_{1}^{\prime\prime}$, data for $H_{ac} = 4$ Oe start to display small deviations with respect to data for $H_{ac} = 1.2$ Oe (not shown). However, only data for $H_{ac} = 4$ Oe allow to reliably extend such investigation over three orders of magnitude for $\nu_{m}$ in a reliable way without too high noise level.

\begin{figure}[t!]
\vspace{6.2cm} \includegraphics{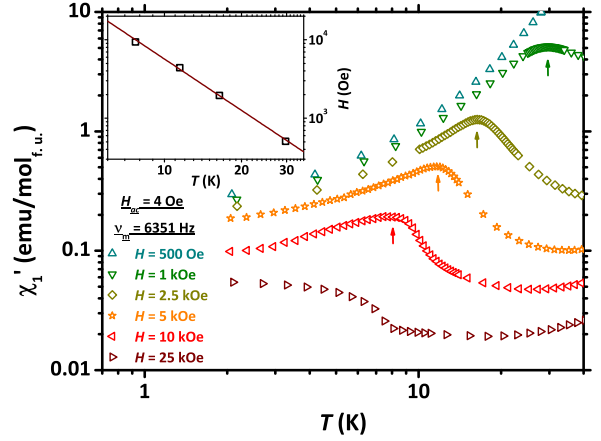}
\caption{\label{FigAt6351HzFirstHarmonicZoomLTAndScaling}(Color online) Main panel: enlargement of $\chi_{1}^{\prime}$ vs. $T$ data already presented in Fig.~\ref{FigAt6351HzFirstHarmonic} evidencing the $H$ dependence of the low-$T$ peak. Inset: $H$ dependence of the temperature values corresponding to the peak of $\chi_{1}^{\prime}$ (see the arrows in the main panel). The continuous lines reproduces the power-law trend reported in Eq.~\eqref{EqPowerLawForTheDependenceOfTheChiPrime}. Values of $H$ on the $y$ axis are reported after correction for the demagnetization factor [see Eq.~\eqref{EqDemagFactMvsH}].}
\end{figure}

Other observations possibly hint to the correctness of the scenario described above concerning the RSG phase at low temperatures. Data reported in Fig.~\ref{FigAt6351HzFirstHarmonicZoomLTAndScaling} show that the low-$T$ anomaly is shifted to lower $T$ values with increasing $H$, similarly to the case of other Sm-based itinerant systems with coexistence of both magnetic $d$ and $f$ electrons.\cite{Oht10} By considering the overall behaviour for $\chi_{1}^{\prime}$ vs. $T$ curves presented in Fig.~\ref{FigAt6351HzFirstHarmonic}, the qualitative similarity with the results reported for AmFA like (Fe$_{1-x}$Mn$_{x}$)$_{75}$P$_{16}$B$_{6}$Al$_{3}$ is evident,\cite{Ber95} in agreement with the data presented above from the other experimental techniques. Experimental points reported in the inset of Fig.~\ref{FigAt6351HzFirstHarmonicZoomLTAndScaling} were analyzed as if they were delimiting the so-called De Almeida-Thouless ($AT$) line, accordingly. After denoting the $T$ values corresponding to the maxima as $T_{AT}(H)$, such line may be defined as\cite{Ber95,Cha11}
\begin{equation}\label{EqDeAlmeidaThouless}
\frac{T_{AT}(0)-T_{AT}(H)}{T_{AT}(0)} \propto H^{2/3}.
\end{equation}
This is the very same behaviour that is typically reported also for the melting line between liquid and glassy phases for vortices in high-$T_{c}$ superconductors.\cite{Pra11,Pra12,Pra13c} However, in the current case of Sm$_{2}$Mo$_{2}$O$_{7}$ no choice of $T_{AT}(0)$ helps to recover a De Almeida - Thouless behaviour.\cite{Ber95} Rather, the $H$ dependence of the maxima in $\chi_{1}^{\prime}$ are well described
\begin{equation}\label{EqPowerLawForTheDependenceOfTheChiPrime}
T \propto H^{-\zeta}
\end{equation}
where for the phenomenological exponent one has $\zeta \simeq 0.45$, well below the expected $\nicefrac{2}{3}$ value. The situation is extremely similar to what was reported for (Fe$_{1-x}$Mn$_{x}$)$_{75}$P$_{16}$B$_{6}$Al$_{3}$, where this apparent anomaly was explained in terms of additional thermally-activated blocking processes.\cite{Ber95} This is also confirmed in the current case of Sm$_{2}$Mo$_{2}$O$_{7}$ by the lack of a clearly divergent contribution to non-linear susceptibility in the low-$T$ region (see Fig.~\ref{FigsThirdHarmonic}). One should consider that the presence of magnetic moments from Sm$^{3+}$ ions in Sm$_{2}$Mo$_{2}$O$_{7}$ may partially influence the detailed features of the RSG phase. However, this aspects complicate the theoretical framework and ask for further more detailed investigations.

\section{Summarizing remarks and conclusions}

\begin{figure}[t!]
\vspace{6.2cm} \includegraphics{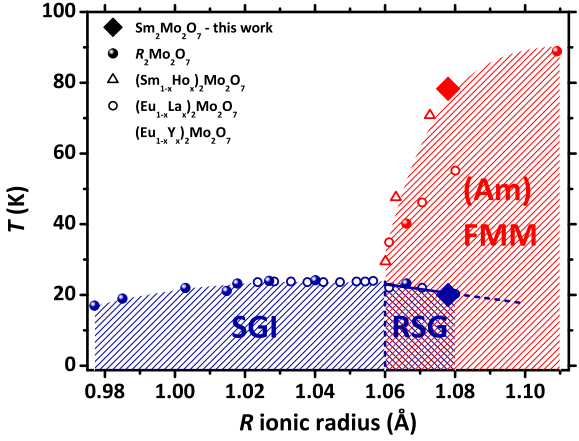}
\caption{\label{FigPhaseDiagram}(Color online) Summarizing electronic phase diagram for pyrochlore molybdates. Results for Sm$_{2}$Mo$_{2}$O$_{7}$ (current work) are displayed by filled diamonds. The red point is $T_{C}^{M}$ while the blue point is $T_{_{RSG}}$ (estimate by $\mu^{+}$SR). Other experimental points are reproduced from Refs.~\onlinecite{Igu09}, \onlinecite{Han07} and \onlinecite{Cla13}. Filled symbols refer to unsubstituted $R$ ions (from left to right: Lu, Yb, Er, Ho, Y, Dy, Tb, Eu, Nd) while empty symbols refer to the substitutions specified in the legend. The case $R$ = Gd is not reported due to the contrasting results in the literature (see, e.~g., Refs.~\onlinecite{Igu09} and \onlinecite{Han07} as opposite to what discussed in Ref.~\onlinecite{Mir06}). The meaning of the labels is explained throughout the text (``Am'' standing for ``amorphous'').}
\end{figure}

We reported on a detailed investigation of a high-quality single crystal of Sm$_{2}$Mo$_{2}$O$_{7}$. The magnetic phase of the Mo$^{4+}$ sublattice clearly displays disordered magnetic features that can be hardly reconciled with the commonly accepted itinerant ferromagnetic state of this material. Accordingly, a new electronic phase diagram for pyrochlore molybdates is proposed in Fig.~\ref{FigPhaseDiagram}, where experimental points for Sm$_{2}$Mo$_{2}$O$_{7}$ ($T_{C}^{M}$ and $T_{_{RSG}}^{\mu}$, see text) are complemented by other materials (data points taken from Refs.~\onlinecite{Igu09}, \onlinecite{Han07} and \onlinecite{Cla13}). 

As main conclusions, the ferromagnetic phase arising within the Mo$^{4+}$ sublattice below $T_{C}^{M}$ is not conventional at all and displays several analogies with amorphous ferromagnetic alloys. This is confirmed by different independent experimental techniques (dc magnetometry, $\mu^{+}$SR and magnetic ac susceptibility). In particular, the scaling analysis of dc magnetometry evidences anomalously high values for the critical exponents, as proper of amorphous magnets rather than conventional ferromagnets. These values are confirmed by the analysis of $\mu^{+}$SR and magnetic ac susceptibility data. At the same time, $\mu^{+}$SR also enlightens the lack of a well-defined long-range magnetic phase reflected in the lack of coherent oscillations of the time-dependence of the $\mu^{+}$ spin polarization. The glassy properties of this disordered magnetic phase are also evidenced by the critical divergence of high-harmonics ac susceptibility. Finally, as typical for several amorphous ferromagnets, a re-entrant spin-glass phase is evidenced at low temperatures by means of both the longitudinal magnetic relaxation of $\mu^{+}$ and by magnetic ac susceptibility. Overall, our results shed new light on the magnetic properties of Sm$_{2}$Mo$_{2}$O$_{7}$ and, accordingly, on the overall electronic phase diagram commonly accepted for pyrochlore molybdates.

\section*{Acknowledgements}
Useful discussions with M. J. P. Gingras and the support of H. Luetkens during $\mu^{+}$SR measurements are gratefully acknowledged. G. Prando acknowledges support by the Humboldt Research Fellowship for Postdoctoral researchers.

\appendix

\section{Correction of demagnetization effects for high-harmonics ac susceptibility data}\label{AppendixDemagnetization}

The output of ac susceptometry measurements is the $t$-dependent magnetization $M_{ac}(t)$ induced in the examined material by a small alternating magnetic field 
\begin{equation}
H_{ac}(t) = H_{ac} e^{\imath\omega t}.
\end{equation}
Measurements can be performed with a polarizing static magnetic field superimposed to $H_{ac}(t)$. Without loss of generality, only the alternating components are considered from now on for both $M_{ac}$ and $H_{ac}$, and the subscripts $ac$ are dropped for the sake of clarity.

Interesting physical insights in the physics of the investigated compound can be obtained by analyzing the discrete Fourier transform of $M(t)$. In particular, one can write
\begin{equation}\label{EqTimeDependentMagnetization}
M(t) = \sum_{n = 1}^{+\infty} M_{n} e^{\imath n \omega t},
\end{equation}
where the complex coefficients $M_{n}$ are straightforwardly defined as
\begin{eqnarray}\label{EqFourierSeriesMagnetization}
M_{n} & \equiv & M_{n}^{\prime} - \imath M_{n}^{\prime\prime} {}\nonumber\\ & = & \frac{1}{2\pi} \int_{0}^{2\pi} M(t) e^{-\imath n \omega t} d\left(\omega t\right).
\end{eqnarray}
From the expression above, the two in-phase and out-of-phase components $M_{n}^{\prime}$ and $M_{n}^{\prime\prime}$ (respectively) are defined for the $n$th-harmonic magnetization. The intrinsic $n$th-harmonic complex ac susceptibilities $\chi_{n}$ may be defined as a function of the intrinsic alternating magnetic field $H_{i}(t) = H_{i} e^{\imath\omega t}$ as
\begin{eqnarray}
M(t) & = & \sum_{n = 1}^{+\infty} M_{n} e^{\imath n \omega t} = \sum_{n = 1}^{+\infty} \frac{M_{n}}{\left(H_{i}\right)^{n}} \left(H_{i} e^{\imath\omega t}\right)^{n} {}\nonumber\\ & = & \sum_{n = 1}^{+\infty} \chi_{n} \left[H_{i}(t)\right]^{n}
\end{eqnarray}
by defining
\begin{equation}\label{EqHarmonics}
\chi_{n} = \frac{M_{n}}{\left(H_{i}\right)^{n}}.
\end{equation}
Experimentally, the ac susceptometer allows one to monitor the behaviour of the complex $M_{n}$ up to a selected value $n^{*}$. Accordingly, the $n$th-harmonic magnetic ac susceptibility $\chi_{n}$ can be calculated from Eq.~\eqref{EqFourierSeriesMagnetization} once the complex quantity $H_{i}$ is known. It is important to stress that the amplitude of the magnetic field $H_{i}$ is in general not the same as $H_{a}$ (experimentally applied field) due to demagnetization effects quantified by the demagnetization factor $D_{m}$ (see Section~\ref{SectionExperimentalsDcMagnetometry}). In particular, the corrected magnetic field should be written as
\begin{equation}\label{EqMagneticFieldCorrected}
H_{i}(t) = H_{a}(t) - 4 \pi D_{m} M(t),
\end{equation}
where $M(t)$ is given by Eq.~\eqref{EqTimeDependentMagnetization}, and substituted into Eq.~\eqref{EqHarmonics} in turn (f.u. molar units are assumed, see Section~\ref{SectionExperimentalsDcMagnetometry}). Besides the $t$-dependent complex exponentials $e^{\imath n \omega t}$, all the quantities in Eq.~\eqref{EqMagneticFieldCorrected} are complex, with the only exception of $H_{a}$ whose imaginary component is zero by definition. The substitution of Eq.~\eqref{EqMagneticFieldCorrected} into Eq.~\eqref{EqHarmonics} leads to the desired expression for $\chi_{n}$ as a function of $H_{a}$. In general, this expression is quite complicate and cumbersome to resolve and some approximation is needed. On the other hand, the limiting case of $n = 1$ corresponds to the standard correction for the demagnetization effects and it is straightforward to be solved exactly.

\subsection{Measurements up to the fundamental harmonic ($n^{*} = 1$)}

Let us focus on the case $n = 1$ first. From Eq.~\eqref{EqTimeDependentMagnetization} one has
\begin{equation}
M(t) = M_{1} e^{\imath \omega t}
\end{equation}
and, by use of Eq.~\eqref{EqMagneticFieldCorrected} (simplifying the $t$ dependent complex exponentials), one gets the expression
\begin{equation}
H_{i} = H_{a} - 4 \pi D_{m} M_{1}.
\end{equation}
Eq.~\eqref{EqHarmonics} is rewritten as
\begin{equation}
M_{1} = \chi_{1} H_{i}
\end{equation}
leading to the equivalent expressions
\begin{eqnarray}\label{EqStandardCorrection}
\frac{M_{1}}{H_{a}} & = & \frac{\chi_{1}}{1 + 4 \pi D_{m} \chi_{1}},\nonumber \\ \chi_{1} & = & \frac{M_{1}/H_{a}}{1 - \left(4 \pi D_{m} M_{1}/H_{a}\right)}.
\end{eqnarray}
By making the real and imaginary components explicit, one has
\begin{equation}
\chi_{1}^{\prime} - \imath \chi_{1}^{\prime\prime} = \frac{\left(M_{1}^{\prime} - \imath M_{1}^{\prime\prime}\right)/H_{a}}{1 - \left[\left(4 \pi D_{m}/H_{a}\right) \left(M_{1}^{\prime} - \imath M_{1}^{\prime\prime}\right)\right]}.
\end{equation}
Accordingly, the real and imaginary components are distinguished as\cite{Mat01,Pra13c}
\begin{eqnarray}\label{EqCaseFirstHarmonic}
\chi_{1}^{\prime} & = & \frac{\left(M_{1}^{\prime}/H_{a}\right) - 4 \pi D_{m}\left[\left(M_{1}^{\prime}/H_{a}\right)^{2} + \left(M_{1}^{\prime\prime}/H_{a}\right)^{2}\right]}{\left[1 - \left(4 \pi D_{m} M_{1}^{\prime}/H_{a}\right)\right]^{2} + \left(4 \pi D_{m} M_{1}^{\prime\prime}/H_{a}\right)^{2}},\nonumber\\
\chi_{1}^{\prime\prime} & = & \frac{\left(M_{1}^{\prime\prime}/H_{a}\right)}{\left[1 - \left(4 \pi D_{m} M_{1}^{\prime}/H_{a}\right)\right]^{2} + \left(4 \pi D_{m} M_{1}^{\prime\prime}/H_{a}\right)^{2}}.\nonumber\\ 
\end{eqnarray}

\subsection{Measurements up to the third harmonic ($n^{*} = 3$)}

Let us now consider the general case beyond the fundamental harmonic. Typically, measurements are discussed in the literature up to $n^{*} = 3$ and for this reason the third harmonic will be considered here as the highest measured one. Coming back to Eqs.~\eqref{EqTimeDependentMagnetization} and \eqref{EqMagneticFieldCorrected}, one has
\begin{equation}\label{EqMagneticFieldCorrectedAgain}
H_{i}(t) = H_{a}(t) - 4 \pi D_{m} M(t)
\end{equation}
where
\begin{equation}\label{EqHarmonicsUpToThree}
M(t) = \sum_{n = 1}^{3} M_{n} e^{\imath n \omega t}.
\end{equation}
Inserting Eq.~\eqref{EqHarmonicsUpToThree} into Eq.~\eqref{EqMagneticFieldCorrectedAgain} would clearly result into cumbersome calculations. As simplifying assumption for the demagnetization correction, one can assume that the expression for $M(t)$ to be substituted into Eq.~\eqref{EqMagneticFieldCorrectedAgain} is the same as considered for the case $n^{*} = 1$, namely
\begin{equation}
M(t) \simeq M_{1} e^{\imath \omega t}.
\end{equation}
This assumption is performed only for the correction to $H_{a}(t)$ by considering that typically $\chi_{1}$ is the dominant contribution if compared to the amplitudes of $\chi_{2}$ and $\chi_{3}$. Analogously to the $n^{*} = 1$ case, after simplifying the $t$ dependent complex exponentials one gets
\begin{equation}
H_{i} \simeq H_{a} - 4 \pi D_{m} M_{1}.
\end{equation}
By considering $n = 1$ in Eq.~\eqref{EqHarmonics}, one has
\begin{equation}\label{EqIntrinsicFieldForComputationOfTheHighHarmonicSusceptibilities}
H_{i} = \frac{H_{a}}{1 + 4 \pi D_{m} \chi_{1}}
\end{equation}
leading to the expressions already reported in Eq.~\eqref{EqCaseFirstHarmonic}. At the same time, Eq.~\eqref{EqIntrinsicFieldForComputationOfTheHighHarmonicSusceptibilities} allows to consider the cases $n = 2$ and $n = 3$ separately. Namely, again from Eq.~\eqref{EqHarmonics} one gets
\begin{eqnarray}
M_{2} & = & \frac{\chi_{2}}{\left(1 + 4 \pi D_{m} \chi_{1}\right)^{2}} H_{a}^{2},\nonumber\\
M_{3} & = & \frac{\chi_{3}}{\left(1 + 4 \pi D_{m} \chi_{1}\right)^{3}} H_{a}^{3}.
\end{eqnarray}
Both these formulas can be conveniently rearranged in order to express the susceptibilities $\chi_{2}$ and $\chi_{3}$ as a function of the observable quantities as
\begin{eqnarray}\label{EqExpressionForHarmonics}
\chi_{2} & = & \frac{M_{2}}{H_{a}^{2}} \left(1 + 4 \pi D_{m} \chi_{1}\right)^{2},\nonumber\\
\chi_{3} & = & \frac{M_{3}}{H_{a}^{3}} \left(1 + 4 \pi D_{m} \chi_{1}\right)^{3}.
\end{eqnarray}

The real and imaginary components of $\chi_{2}$ and $\chi_{3}$ should be put in explicit form as in the case of Eq.~\eqref{EqCaseFirstHarmonic}. Starting from Eq.~\eqref{EqCaseFirstHarmonic} one obtains
\begin{eqnarray}\label{EqCaseSecondHarmonic}
\chi_{2}^{\prime} & = & \left(M_{2}^{\prime}/H_{a}^{2}\right) \left[\left(1 +  4 \pi D_{m} \chi_{1}^{\prime}\right)^{2} - \left(4 \pi D_{m} \chi_{1}^{\prime\prime}\right)^{2}\right]\nonumber \\ && - 8 \pi D_{m} \chi_{1}^{\prime\prime} \left(M_{2}^{\prime\prime}/H_{a}^{2}\right) \left(1 +  4 \pi D_{m} \chi_{1}^{\prime}\right) \nonumber\\
\chi_{2}^{\prime\prime} & = & 8 \pi D_{m} \chi_{1}^{\prime\prime} \left(M_{2}^{\prime}/H_{a}^{2}\right) \left(1 +  4 \pi D_{m} \chi_{1}^{\prime}\right)\nonumber \\ && + \left(M_{2}^{\prime\prime}/H_{a}^{2}\right) \left[\left(1 +  4 \pi D_{m} \chi_{1}^{\prime}\right)^{2} - \left(4 \pi D_{m} \chi_{1}^{\prime\prime}\right)^{2}\right]\nonumber \\
\end{eqnarray}
for the case $n = 2$ while for $n = 3$ one gets
\begin{eqnarray}\label{EqCaseThirdHarmonic}
\chi_{3}^{\prime} & = & \left(M_{3}^{\prime}/H_{a}^{3}\right) \left[\left(1 +  4 \pi D_{m} \chi_{1}^{\prime}\right)^{3}\right.\nonumber \\ && \left.- 3 \left(1 +  4 \pi D_{m} \chi_{1}^{\prime}\right) \left(4 \pi D_{m} \chi_{1}^{\prime\prime}\right)^{2}\right] - \left(M_{3}^{\prime\prime}/H_{a}^{3}\right)\nonumber \\ && \times \left[12 \pi D_{m} \chi_{1}^{\prime\prime} \left(1 +  4 \pi D_{m} \chi_{1}^{\prime}\right)^{2} - \left(4 \pi D_{m} \chi_{1}^{\prime\prime}\right)^{3}\right]\nonumber\\
\chi_{3}^{\prime\prime} & = & \left(M_{3}^{\prime}/H_{a}^{3}\right)\nonumber \left[12 \pi D_{m} \chi_{1}^{\prime\prime} \left(1 +  4 \pi D_{m} \chi_{1}^{\prime}\right)^{2}\right. \\ && \left. - \left(4 \pi D_{m} \chi_{1}^{\prime\prime}\right)^{3}\right] + \left(M_{3}^{\prime\prime}/H_{a}^{3}\right) \left[\left(1 +  4 \pi D_{m} \chi_{1}^{\prime}\right)^{3}\right. \nonumber \\ && \left. - 3 \left(1 +  4 \pi D_{m} \chi_{1}^{\prime}\right) \left(4 \pi D_{m} \chi_{1}^{\prime\prime}\right)^{2}\right].\nonumber \\
\end{eqnarray}




\end{document}